\documentclass[reprint, amsmath,amssymb, aps, prl, superscriptaddress,nobibnotes]{revtex4-2}

\usepackage{graphicx}
\usepackage{dcolumn}
\usepackage{bm}
\usepackage{multirow}
\usepackage{amsfonts}
\usepackage{amssymb}
\usepackage{amsthm}
\usepackage{array}
\usepackage{amsmath}
\usepackage{verbatim}
\usepackage{hyperref}
\usepackage{color}
\usepackage{bbold}
\usepackage{epstopdf}
\usepackage{mathtools}
\usepackage{MnSymbol}
\usepackage{tikz}
\usepackage[export]{adjustbox}
\usepackage{subfigure}
\usepackage[normalem]{ulem}

\usepackage{lipsum}

\usepackage{xcolor}
\usepackage{ulem}

\newcommand{\affa}{State Key Laboratory of Low Dimensional Quantum Physics, Department of Physics, Tsinghua University, Beijing 100084, China}
\newcommand{\affb}{Beijing Academy of Quantum Information Sciences, Beijing 100193, China}
\newcommand{\affc}{Korea Institute for Advanced Study, Seoul 02455, Korea}
\newcommand{\affd}{Blackett Laboratory, Imperial College London, London SW7 2AZ, United Kingdom}
\newcommand{\affe}{Hefei National Laboratory, Hefei 230088, P. R. China}
\newcommand{\afff}{Frontier Science Center for Quantum Information, Beijing 100084, China}
\newcommand{\affg}{Institute for Quantum Science and Technology, College of Science, National University of Defense Technology, Changsha 410073, China}
\newcommand{\affh}{Hunan Key Laboratory of Mechanism and Technology of Quantum Information, Changsha 410073, China}

\newcommand{\Yb}{$^{171}\rm{Yb}^+$}
\newcommand{\Ba}{$^{138}\rm{Ba}^+$}

\newcommand{\del}[1]{\iffalse #1\fi}

\theoremstyle{remark}
\newtheorem{proposition}{Proposition}

\newtheorem{theorem}{Theorem}

\newtheorem{observation}{Observation}

\newcommand{\ket}[1]{\left\vert#1\right\rangle}
\newcommand{\bra}[1]{\left\langle#1\right\vert}

\def\bra#1{\langle #1|}
\def\ket#1{\left|#1 \right>}

\def\Tr{\mbox{Tr}}

\begin{document}

\title{Snapshotting Quantum Dynamics at Multiple Time Points}

\author{Pengfei Wang}
\thanks{,$^\dagger$ First three authors contributed equally.}
\address{\affb}
\address{\affa}
\author{Hyukjoon Kwon}
\email{hjkwon@kias.re.kr}
\address{\affc}
\author{Chun-Yang Luan}
\thanks{,$^\dagger$ First three authors contributed equally.}
\address{\affa}
\address{\affg}
\address{\affh}
\author{Wentao Chen}
\address{\affa}
\author{Mu Qiao}
\address{\affa}
\author{Zinan Zhou}
\address{\affa}
\author{Kaizhao Wang}
\address{\affa}
\author{M. S. Kim}
\email{m.kim@imperial.ac.uk}
\address{\affc}
\address{\affd}
\author{Kihwan Kim}
\email{kimkihwan@mail.tsinghua.edu.cn}
\address{\affb}
\address{\affa}
\address{\affe}
\address{\afff}

\begin{abstract}
Measurement-induced state disturbance is a major challenge in obtaining quantum statistics at multiple time points. We propose a method to extract dynamic information from a quantum system at intermediate time points, namely snapshotting quantum dynamics. 
To this end, we apply classical post-processing after performing the ancilla-assisted measurements to cancel out the impact of the measurements at each time point. Based on this, we reconstruct a multi-time quasi-probability distribution (QPD) that correctly recovers the probability distributions at the respective time points.
Our approach can also be applied to simultaneously extract exponentially many correlation functions with various time-orderings. We provide a proof-of-principle experimental demonstration of the proposed protocol using a dual-species trapped-ion system by employing \Yb~and \Ba~ions as the system and the ancilla, respectively. Multi-time measurements are performed by repeated initialization and detection of the ancilla state without directly measuring the system state. The two- and three-time QPDs and correlation functions are reconstructed reliably from the experiment, negativity and complex values in the QPDs clearly indicate a contribution of the quantum coherence throughout dynamics.
\end{abstract}
\maketitle

\section{Introduction}

A striking difference between quantum mechanics and classical mechanics arises from understanding the measurements. In quantum mechanics, the uncertainty principle asserts that it is impossible to define a joint probability distribution of statistical properties of non-commuting variables, thus prohibiting a description of quantum physics using classical probability theory. This leads to the introduction of quasi-probability distributions (QPDs), a prototypical example of which is the Wigner function~\cite{Wigner32} describing quantum phase space. Another important class of QPDs is the Kirkwood-Dirac (KD) distribution~\cite{Kirkwood33, Dirac45,arvidssonshukur2024}, which can be applied to any two incompatible sets of measurement operators. The nonclassical features in these QPDs, characterized by negative~\cite{Wigner32, Margenau61} or even non-real values~\cite{Kirkwood33, Dirac45}, have been investigated within the realms of quantum foundations~\cite{Bell66, Leggett85}, closely connected to quantum contextuality~\cite{spekkens2008negativity, ferrie2008frame, hofmann2011role, de2021complete, hance2023contextuality, Wagner2023simple},
and recently recognized as a resource in quantum computing~\cite{veitch2012negative, mari2012positive, Pusey2014Anomalous, howard2014contextuality, pashayan2015estimating, rahimi2016sufficient} and quantum metrology~\cite{Kwon19, Arvidsson-Shukur2020, lostaglio2020certifying, lupu2022negative}.

The same principle is applied when performing sequential measurements during the evolution of a quantum state. The double-slit experiment serves as an illustration of this phenomenon: attempting to extract path information causes the final interference patterns to disappear. Consequently, one cannot obtain a classical joint probability distribution that simultaneously describes both the which-path information and the final position of the particle. The absence of the classical probability description of quantum mechanical processes gives rise to the nonclassicality of temporal correlation described by the Leggett-Garg inequality~\cite{Leggett85} and the no-go theorem for defining work observables in quantum thermodynamics~\cite{perarnau2017no}. Meanwhile, with recent advances in quantum information science, there has been an increasing demand to study multi-time quantum statistics to explore exotic features of quantum dynamics, such as information spreading throughout quantum dynamics~\cite{Maldacena16, Landsman19}. Recently, it has also been shown that monitoring the dynamics of a quantum system at multiple time points can witness entanglement~\cite{Jayachandran23}.

On the other hand, a major challenge arises when attempting to access these quantum correlations over time in experiments. As observed from the double-slit experiment, direct measurements performed on a quantum state wash out quantum coherences so that they can no longer contribute to the subsequent dynamics. The destructive and irreversible effect of measurements on a quantum system raises an ongoing question: Is it possible to extract information of the system at intermediate points in time throughout quantum dynamics while minimizing the impact of the measurement on subsequent events? The most widely adopted method is the use of weak measurements~\cite{Aharonov88, Leggett89, Brun02, resch2004extracting, mitchison2007sequential, wiseman_milburn_2009} (see Refs.~\cite{lundeen2011direct, lundeen2012procedure, bamber2014observing, piacentini2016measuring, thekkadath2016direct, kim2018direct} for experimental realizations) to gain little information with little disturbance of the system~\cite{Fuchs96, myrvold2009no, Buscemi14}.
Over the years, various quantum measurement schemes~\cite{johansen2007quantum, resch2004extracting, mitchison2007sequential} beyond weak measurement have been proposed to extract temporal quantum correlation functions. Alternative approaches have also been explored, including those utilizing long-time entanglement between the system and the ancilla~\cite{mazzola2013measuring, pedernales2014efficient, swingle2016measuring, halpern2017jarzynski}, as well as methods involving multiple copies of quantum states for each trial~\cite{yao2016interferometric, perarnau2017no, wu2019experimentally}.

\begin{figure}
    \centering
    \includegraphics[width=88mm]{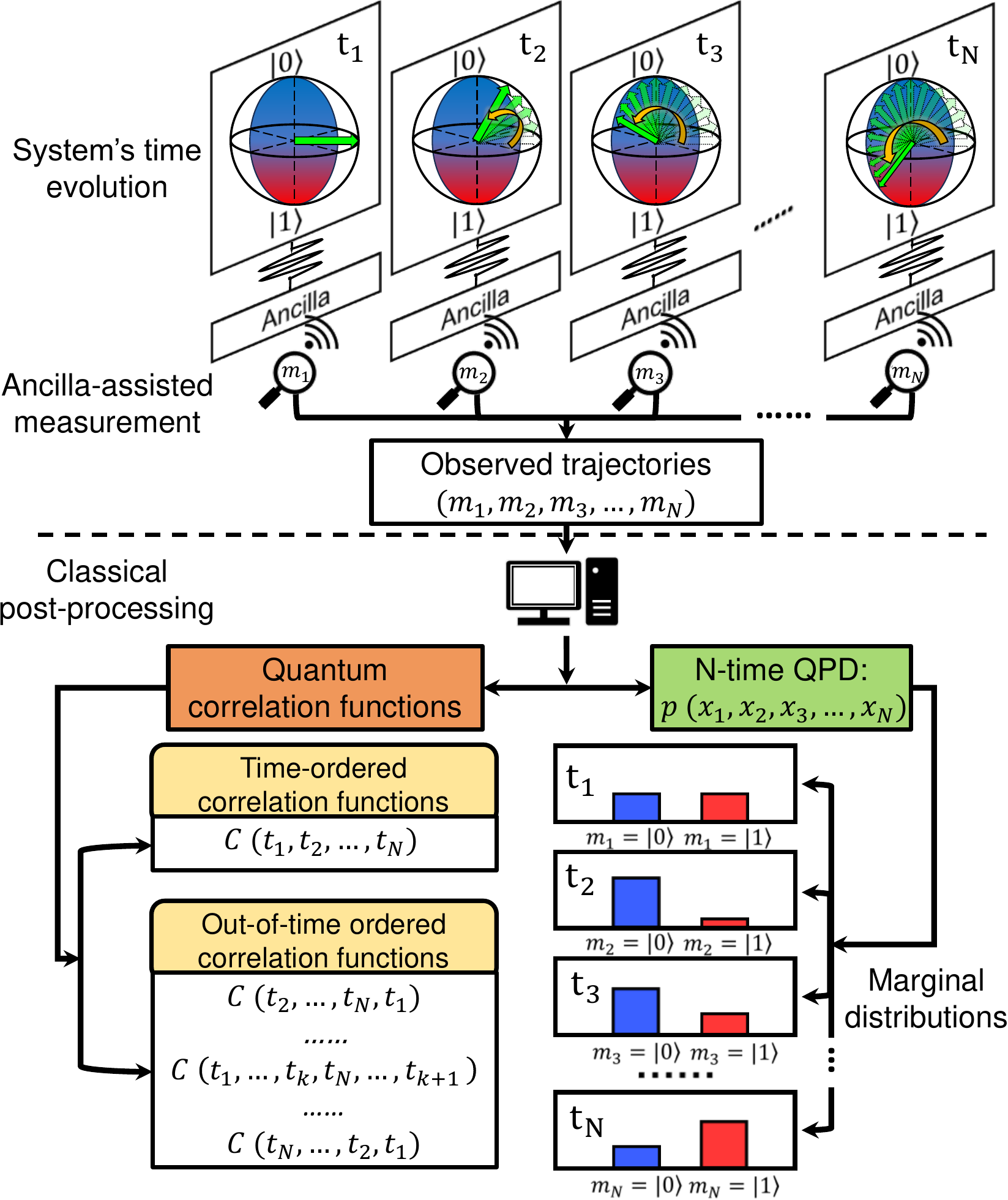}
    \caption{Schematic procedure for snapshotting quantum dynamics. 
    Various types of information on quantum dynamics are obtained simultaneously through classical post-processing of the intermediate measurement outcomes. These include the multi-time  quasi-probability distribution (QPD) with the correct marginal probabilities at the respective time points, as well as both time-ordered and out-of-time-ordered correlation functions.
    }
    \label{fig:dynamics snapshotting}
\end{figure}

In this work, we propose a method to simultaneously extract multiple types of quantum system's dynamical information through ancilla-assisted measurements at intermediate time points, which we term snapshotting quantum dynamics (see Fig.~\ref{fig:dynamics snapshotting}). The key idea is to cancel out the impact of the measurements via classical post-processing of the outcomes, which enables us to obtain multi-time QPDs as well as both time-ordered and out-of-time-ordered quantum correlation functions. We experimentally demonstrate the proposed method in a \Yb-\Ba~trapped-ion system by reconstructing QPD and quantum correlation functions with the full contribution of coherence up to three-time points.

\section{Results}
\subsection{QPD for multiple time points} 
Suppose a quantum state is $\rho_{t_i}$ at $t_i$ and evolves in time. The quantum state at a certain time $t_j$ can be expressed as 
\begin{equation}
\rho_{t_j} = {\cal N}_{t_i \rightarrow t_j} (\rho_{t_i}),
\end{equation}
where ${\cal N}_{t_i \rightarrow t_j} $ is a completely positive trace-preserving quantum channel describing the evolution from time $t_i$ to $t_j$. To obtain the information of the quantum state at time $t_i$, one may perform measurements given by a set of projection operators $\Pi_{x_i} = \ket{x_i}\bra{x_i}$ satisfying $\sum_{x_i} \Pi_{x_i} = \mathbb{1}$, which leads to the outcome distribution of $x_i$ at time $t_i$, $p(x_i;t_i) = \Tr[ \rho_{t_i} \Pi_{x_i}]$. After the measurement is performed, the state collapses to $\ket{x_i}\bra{x_i}$. Such a projective measurement incurs a critical issue when obtaining quantum statistics for more than two sequential time points. For example, the joint distribution of outcomes by performing projective measurements at times $t_1$ and $t_2$ can be written as $p^{\rm proj.}(x_1, x_2; t_1, t_2) = p(x_1;t_1) \Tr[ {\cal N}_{t_1 \rightarrow t_2} (\ket{x_1}\bra{x_1}) \Pi_{x_2}]$.
However, the marginal distribution at time $t_2$ obtained from the joint distribution $p^{\rm proj.}(x_1, x_2; t_1, t_2)$ does not match the statistics without the measurement at time $t_1$, i.e., $\sum_{x_1} p^{\rm proj.}(x_1,x_2; t_1,t_2) \neq p(x_2;t_2)$. 
This invokes the so-called measurement problem that the wave-function collapse induced by the measurement cannot be explained as a direct consequence of the Schr\"{o}dinger equation~\cite{Zurek2003Decoherence, Schlosshauer2005Decoherence, Bassi2013Models}. Consequently, the joint distribution of projective measurement outcomes becomes unsuitable for providing a complete description of quantum dynamics.

To address such a problem, one can introduce a two-time joint distribution,

\begin{equation}
\begin{aligned}
p(x_1, x_2 ; t_1, t_2) &\equiv \Tr[ {\cal N}_{t_1 \rightarrow t_2} (\rho_{t_1} \Pi_{x_1}) \Pi_{x_2}] \\
&= \Tr[ ({\cal M}_{x_2} \circ {\cal N}_{t_1 \rightarrow t_2} \circ {\cal M}_{x_1}) (\rho_{t_1}) ],
\end{aligned}
\end{equation}
where ${\cal M}_x(\rho) \equiv \rho \Pi_x$. We note that $p(x_1, x_2; t_1, t_2)$ is well-normalized, $\sum_{x_1, x_2} p(x_1, x_2;t_1,t_2) = 1$, and correctly indicates the marginal distribution, $\sum_{x_1} p(x_1, x_2 ; t_1, t_2) = p(x_2;t_2)$. However, $p(x_1, x_2;t_1,t_2)$ can have complex values, i.e., being a QPD, and can be understood as the KD distribution~\cite{Kirkwood33, Dirac45} of the two different measurement operators $\Pi_{x_1}$ and ${\cal N}_{t_1 \rightarrow t_2}^\dagger(\Pi_{x_2})$~\cite{lostaglio2023kirkwood}.
Such a distribution has been recently rediscovered to provide a useful mathematical formalism to explore the concept of work and the fluctuation theorems in quantum thermodynamics~\cite{perarnau2017no, halpern2017jarzynski, halpern2018quasiprobability, Allahverdyan04, Kwon19X, lostaglio2023kirkwood,  upadhyaya2023happens, zhang2024quasiprobability}.

The QPD based on the KD distribution was also generalized to multiple-time points~\cite{halpern2018quasiprobability, lupu2022negative}. In this paper, we define $N$-time QPD as 
\begin{equation}
\label{eq:joint_dist}
\begin{aligned}
&p(x_1, x_2, \cdots, x_N ; t_1, t_2, \cdots, t_N) \\
&\equiv \Tr[ ({\cal M}_{x_N} \circ {\cal N}_{t_{N-1} \rightarrow t_N}\circ \cdots \circ {\cal M}_{x_2} \circ {\cal N}_{t_1 \rightarrow t_2} \circ {\cal M}_{x_1}) (\rho_{t_1}) ].
\end{aligned}
\end{equation}
When the quantum state at each time point commutes with the measurement operator, i.e., $[\rho_{t_i}, \Pi_{x_i}]=0$ for all $t_i$, the distribution coincides with the classical joint distribution obtained from sequential projective measurements. 
Consequently, nonclassical values, i.e., negative or non-real values in the QPD capture the coherence of the system state within the measurement basis by witnessing the non-commutativity between the state and the measurement operator (see, e.g., Refs.~\cite{de2021complete, Arvidsson-Shukur_2021, Wagner2023simple, hance2023contextuality} for more detailed analysis). Throughout the manuscript, we will also use a simplified notation $p(x_1, x_2, \cdots, x_N) = p(x_1, x_2, \cdots, x_N; t_1, t_2, \cdots, t_N)$ when the time sequence is trivial from the context.

An important property of the $N$-time QPD is that it correctly reproduces marginal distributions, satisfying 
\begin{equation}
\label{eq:marginal_prob}
\sum_{x_k} p(x_1, \cdots, x_N) = p(x_1, \cdots, x_{k-1}, x_{k+1}, \cdots, x_N)
\end{equation}
for any $k=1,2,\cdots,N$, where
$p(\cdots, x_{k-1}, x_{k+1}, \cdots) \quad \equiv \Tr[ ( \cdots \circ {\cal M}_{x_{k+1}} \circ {\cal N}_{t_{k-1} \rightarrow t_{k+1}}\circ {\cal M}_{x_{k-1}} \circ \cdots ) (\rho_{t_1})]$
is a joint QPD without performing a measurement ${\cal M}_{t_k}$ at time $t_k$. This is known as the Kolmogorov consistency condition in classical probability theory~\cite{chow2012probability}.
In other words, the $N$-time QPD incorporates all the information from the $k$-time QPDs $p(x_{i_1}, x_{i_2}, \cdots, x_{i_k};t_{i_1}, t_{i_2}, \cdots, t_{i_k})$ for any sub-time sequences with $1 \leq i_1 < i_2 < \cdots < i_k \leq N$. 
In particular, the marginal distribution of the QPD at a single time $t_i$, $p(x_i)$, becomes real and non-negative, correctly indicating the probability distribution of the measurement outcome $x_i$ at time $t_i$. 

We also note that the $N$-time QPD cannot simply be expressed as a product of two-time QPDs, since it does not obey the Markov chain property~\cite{chow2012probability}, i.e., 
$p(x_1, \cdots, x_N) \neq p(x_N|x_{N-1}) \cdots p(x_2|x_1) p(x_1)$ with $p(x_k | x_{k-1}) = \frac{p(x_{k-1},x_{k})}{p(x_{k-1})}$.
We highlight that the quantum channel ${\cal N}_{t_k \rightarrow t_{k+1}}$ for each time interval is Markovian, hence the non-Markovianity arises from the effect of ${\cal M}_{x_k}$.

\subsection{Snapshotting quantum dynamics}
The primary challenge in dealing with QPDs is that experimental reconstruction is not straightforward due to the presence of negative or non-real values. For the $N$-time QPD defined in Eq.~\eqref{eq:joint_dist}, this stems from the fact that ${\cal M}_{x_i}(\rho_{t_i}) = \rho_{t_i} \Pi_{x_i}$ at each time $t_i$ is a non-physical process that does not yield a Hermitian matrix. 
Our key observation to overcome this issue is that ${\cal M}_x$ can be alternatively expressed as a weighted sum of ${\cal K}_m(\rho) \equiv K_m \rho K_m^\dagger = p^{\cal K}(m) \rho^{\cal K}_m$, which can be interpreted as the result of a generalized measurement with outcome $m$~\cite{nielsen2002quantum}.
The probability of the outcome is given by $p^{\cal K}(m) = {\rm Tr}[{\cal K}_m(\rho)] = {\rm Tr}[\rho K_m^\dagger K_m]$, and the state after the measurement becomes $\rho^{\cal K}_m = \frac{K_m \rho K_m^\dagger}{p^{\cal K}(m)}$. The measurement operators compose a set of Kraus operators $\{ K_m \}$, satisfying $\sum_m K_m^\dagger K_m = \mathbb{1}$.

\begin{figure}
    \centering
    \includegraphics[width=88mm]{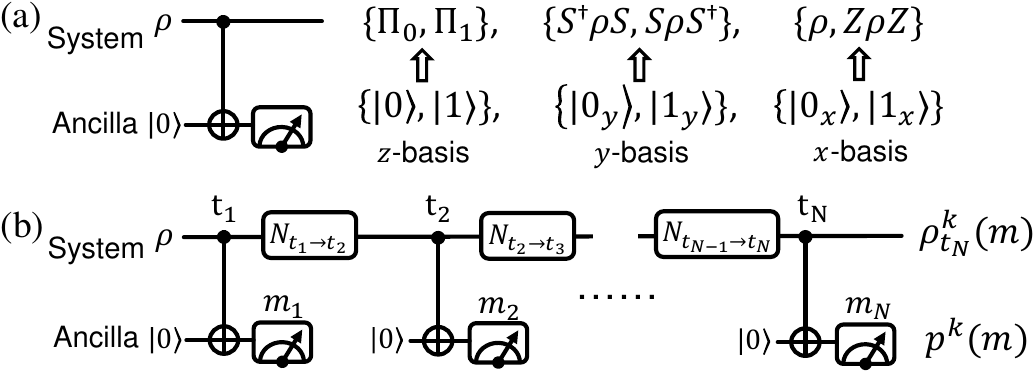}
    \caption{
    Quantum circuit for ancilla-assisted measurement. (a) Ancilla-assisted measurement for realizing Kraus operators. By performing $z$-, $y$- and $x$-basis measurements on the ancilla, the system state is updated depending on the measurement outcome.
    (b) The quantum circuit to obtain the QPD $p(x_1, \cdots, x_N)$ for a qubit system. ${\cal N}_{t_{N-1} \rightarrow t_{N}}$ describes the dynamics of the system from $t_{N-1}$ to $t_{N}$. The ancilla-assisted measurement is performed at each time $t_i$ with the outcome $m_i$. The system state is updated to $\rho_{t_N}^{\cal K} (m_1, \cdots, m_N)$ when the measurement outcomes read $(m_1, \cdots, m_N)$, which happens with probability $p^{\cal K}(m_1, \cdots, m_N)$.}
    \label{fig:concept}
\end{figure}

More explicitly, 
we aim to prove the following expression,
\begin{equation} 
\label{eq:decomp}
{\cal M}_{x} (\rho) = \sum_{m} \gamma_{x m} {\cal K}_{m} (\rho),
\end{equation}
with complex-valued coefficients $\gamma_{xm}$. 
The complex coefficients can be implemented via classical post-processing by weighting the measurement outcomes differently, which will be discussed in more detail.
For example, the projection onto the computational basis of a qubit system, $\Pi_x = \ket{x}\bra{x}$ with $x=0,1$, can be decomposed into 
\begin{equation}
\label{eq:qubit_decomp}
\rho \Pi_x = \frac{\rho - Z \rho Z + 4 \Pi_x \rho \Pi_x - i (-1)^x S \rho S^\dagger + i (-1)^x S^\dagger \rho S}{4},
\end{equation}
where $S = \ket{0}\bra{0} + i \ket{1}\bra{1}$ is the phase gate.
We further note that any Kraus operators can be realized by ancilla-assisted measurements~\cite{nielsen2002quantum}. For the qubit case in Eq.~\eqref{eq:qubit_decomp}, the CNOT gate between the system and the ancilla followed by the ancilla measurement in the $x$-, $y$-, and $z$-bases leads to the set of Kraus operators, $\{ K_m\} = \left\{ \frac{\Pi_0}{\sqrt{3}}, \frac{\Pi_1}{\sqrt{3}}, \frac{S^\dagger}{\sqrt{6}}, \frac{S}{\sqrt{6}}, \frac{\mathbb{1}}{\sqrt{6}}, \frac{Z}{\sqrt{6}}  \right\}$~(see Fig.~\ref{fig:concept}(a)).

This observation can be further generalized to any $d$ dimensional quantum system as follows:
\begin{theorem} 
\label{thm:Kraus_decomp}
For any set of projectors $\{ \Pi_x \}_{x=0}^{d-1}$ acting on a $d$-dimensional quantum state $\rho$, one can always construct a set of Kraus operators $\{K_m\}$ from ancilla-assisted measurement and find coefficients $\gamma_{xm}$ satisfying Eq.~\eqref{eq:decomp}. The Kraus operators are determined by the informationally complete measurement on a $d$-dimensional ancilla state after its interaction with the system.
\end{theorem}
The proof of Theorem~\ref{thm:Kraus_decomp} with more details can be found in Methods.

Now, we introduce a protocol for snapshotting quantum dynamics via intermediate measurements. As an illustrative example, we show that the two-time QPD $p(x_1, x_2; t_1, t_2)$ can be obtained by ancilla-assisted measurements at times $t_1$ and $t_2$ as follows. 
At time $t_1$, we interact the state $\rho_{t_1}$ with the ancilla and measure the ancilla state.
For the ancilla state's outcome $m_1$, the system state is updated to $\rho^{\cal K}_{t_1}(m_1) = \frac{ {\cal K}_{m_1} (\rho_{t_1})}{p^{\cal K}_{t_1}(m_1)}$ with probability $p^{\cal K}_{t_1}(m_1) = {\rm Tr} [{\cal K}_{m_1} (\rho_{t_1})]$. Subsequently, the system evolves to $\rho^{\cal K}_{t_2}(m_1)  = {\cal N}_{t_1 \rightarrow t_2} \left(\rho^{\cal K}_{t_1}(m_1) \right) $ from time $t_1$ to $t_2$. We then perform the second measurement at time $t_2$. When the outcome is $m_2$, the system state is updated to $\rho^{\cal K}_{t_2}(m_1, m_2) = \frac{{\cal K}_{m_2} (\rho^{\cal K}_{t_2}(m_1))}{p^{\cal K}_{t_2}(m_2|m_1)}$ with conditional probability $p^{\cal K}_{t_2}(m_2 | m_1) = {\rm Tr}[{\cal K}_{m_2} (\rho^{\cal K}_{t_2}(m_1))]$ for a given first measurement outcome $m_1$. The joint probability of the sequential measurement outcome $(m_1, m_2)$ becomes $p^{\cal K}(m_1, m_2) = p^{\cal K}_{t_1}(m_1) p^{\cal K}_{t_2}(m_2 | m_1)  = \Tr[ ({\cal K}_{m_2} \circ {\cal N}_{t_1 \rightarrow t_2} \circ {\cal K}_{m_1}) (\rho_{t_1})]$. We note that this joint probability is a classical probability distribution that can be obtained directly from the outcome statistics.

We emphasize that the QPD $p(x_1, x_2)$ and the classical joint distribution $p^{\cal K}(m_1, m_2)$ are linked through Eq.~\eqref{eq:decomp} in the form of $p(x_1, x_2) = \sum_{m_1, m_2} \gamma_{x_1 m_1} \gamma_{x_2 m_2} p^{\cal K}(m_1, m_2)$. Therefore once the probability distribution $p^{\cal K}(m_1,m_2)$ is obtained from the measurement outcomes, $p(x_1, x_2)$ can be reconstructed via classical post-processing by the weighted sum of these probabilities.

As shown in Fig.~\ref{fig:concept}(b), it is straightforward to repeat this protocol for multiple time points, which leads to the following expression of the $N$-time QPD:
\begin{equation} 
\label{eq:QPDs_from_traj}
\begin{aligned}
p(x_1, \cdots, x_N) 
&= \sum_{m_1,\cdots,m_N} p^{\cal K} (m_1, \cdots, m_N) \left[ \prod_{i=1}^N \gamma_{x_i m_i} \right]\\
&=\mathbb{E} \left[ \prod_{i=1}^N \gamma_{x_i m_i} \right],
\end{aligned}
\end{equation}
where $\mathbb{E}[\cdot]$ denotes averaging over all possible sequential measurement outcomes $(m_1, \cdots, m_N)$, which can be understood as the observed trajectories of the quantum dynamics, following the distribution $p^{\cal K}(m_1, \cdots, m_N) = \Tr[ ({\cal K}_{m_N} \circ {\cal N}_{t_{N-1} \rightarrow t_N}\circ \cdots \circ {\cal N}_{t_1 \rightarrow t_2} \circ {\cal K}_{m_1}) (\rho_{t_1}) ]$.

We note that the number of observed trajectories to be collected to reconstruct quantum statistics $p(x_1, \cdots, x_N)$ is greater than that for classical statistics $p^{\cal K}(m_1, \cdots, m_N)$ with the same precision. More precisely, the number of trajectories $M_{\rm traj}$ to estimate $p(x_1,  \cdots, x_N)$ for each time point within a fixed precision $\epsilon$ with probability $1-\delta$ scales as $M_{\rm traj} = \frac{2 \left(\max_{x,m}|\gamma_{xm}|\right)^{2N}}{\epsilon} \ln(2/\delta)$ from Hoeffding's inequality~\cite{Hoeffding63}. Our protocol can also be applied to local projectors of multi-qubit systems with the same sampling overhead. We also provide a systematic algorithm to find the optimal coefficient $\gamma_{xm}$ to reconstruct the joint probability with the minimum number of measurement outcomes (see Methods).

Our method shares an advantage with sequential weak measurements~\cite{mitchison2007sequential, piacentini2016measuring,thekkadath2016direct, kim2018direct}, requiring a short interaction time between the system and the ancilla. However, the main difference lies in the ability of our protocol to completely cancel out the measurement effect through classical post-processing without approximating the system state in the weakly interacting regime. The resource requirement of the proposed protocol can be compared to other schemes for obtaining the KD distribution, while explicit comparisons between these protocols are challenging due to their different natures (see also Refs.~\cite{halpern2018quasiprobability, lostaglio2023kirkwood} for an overview). Compared to the two-point measurement-based scheme~\cite{johansen2007quantum}, which requires the measurement of multiple measurement distributions to infer the KD distribution, our protocol only requires a single measurement distribution $p^{\cal K}(m_1, m_2)$. The main difference compared to the weak measurement-based schemes~\cite{resch2004extracting, mitchison2007sequential} is that our protocol does not need to implement a weak coupling between the system and the ancilla to ensure that the system is undisturbed. While the schemes~\cite{buscemi2013direct, calderaro2018direct} entailing strong measurements for two-time points share a similar structure with our protocol, our protocol provides a straightforward multi-time generalization.  Another scheme based on characteristic function estimation~\cite{mazzola2013measuring} requires classical post-processing of the inverse Fourier transform, while our protocol has a relatively simple post-processing with pre-determined coefficients $\gamma_{x_i m_i}$. The interference-based scheme~\cite{halpern2017jarzynski} requires overlapping measurement and tomography of a quantum state in some occasions, both of which are not required in our scheme. A recently proposed quantum circuit model based on the block-encoding~\cite{rall2020quantum} could have a lower sampling cost than our protocol, but its application is limited to unitary dynamics and requires the implementation of the inverse unitary channel.

Taking into account the physical constraints, our protocol has a short time of system-ancilla coherence during a measurement process at each time, which has an advantage over interferometric schemes~\cite{mazzola2013measuring, halpern2017jarzynski} that require a long time of system-ancilla coherence throughout the entire protocol. While applying measurements in sequential times could also be a challenging task, it has been realized on various physical platforms~\cite{souza2011scattering, piacentini2016measuring, xin2017measurement, ringbauer2018multi, wu2019experimentally, del2024robust}.
We also note that such intermediate measurement has been an active research area, as being an essential technique for quantum information processing, such as syndrome detection for quantum error correction~\cite{nielsen2002quantum}.

In the following section, we discuss that the classical post-processing of the sequential measurement outcomes leads to a unique feature of our approach, which allows the simultaneous extraction of exponentially many correlation functions.

\subsection{Extraction of multi-time correlation functions}
While the $N$-time QPD provides valuable information about the marginal distribution at each time, its utility can even go beyond that. We demonstrate that correlation functions with different time-orderings can be obtained simultaneously from the $N$-time QPD. The quantum correlation function of an observable $A$ throughout unitary quantum dynamics given by $U_{t_i \rightarrow t_j}$ is defined as
\begin{equation}
\begin{aligned}
C(t_1, \cdots, t_N) \equiv \langle A(t_1)\cdots A(t_N) \rangle \equiv \Tr[ \rho_{t_0} A(t_1)\cdots A(t_N)],
\end{aligned}
\end{equation}
where $A(t_i) = U_{t_0 \rightarrow t_i}^\dagger A  U_{t_0 \rightarrow t_i}$ is an observable in the Heisenberg picture. If the time sequence is given in increasing order, i.e., $t_1 \leq t_2 \leq \cdots \leq t_N$, the correlation function is called time-ordered, otherwise it is called out-of-time-ordered.

Using the eigenvalue decomposition of the observable, $A = \sum_{x} a_x \Pi_x$, the time-ordered correlation function can be expressed in terms of the QPD as
\begin{equation}
C(t_1, \cdots, t_N) = \sum_{x_1, \cdots, x_N} a_{x_1} \cdots a_{x_N} p(x_1, \cdots, x_N).
\label{eq:QPDandCorrelationFunction}
\end{equation}
Furthermore, as the $N$-time QPD contains any $k$-time QPD with $k\leq N$, all lower order correlation functions can also be obtained from $p(x_1, \cdots, x_N)$. For example, one can simultaneously obtain a complete set of time-ordered correlation functions $\{ C(t_1), C(t_2), C(t_3)\}$, $C(t_1, t_2), C(t_2, t_3), C(t_1, t_3)\}$, and $C(t_1, t_2, t_3)$ from the three-time QPD $p(x_1, x_2, x_3)$.

More surprisingly, the snapshotting method can be utilized to obtain a family of out-of-time-ordered quantum correlation functions, summarized by the following observation.
\begin{observation}
\label{thm:corr_ft_out}
All correlation functions $C(t_{\mu_1}, \cdots, t_{\mu_j}, t_{\mu_{j+1}} \cdots, t_{\mu_k})$ with $\mu_1 \leq \mu_2 \leq \cdots \leq \mu_{j-1} \leq \mu_j$ and $\mu_j \geq \mu_{j+1} \geq \cdots \geq \mu_{k-1} \geq \mu_k$ for some $\mu_j \leq N$ can be simultaneously deduced from the distribution of observed trajectories $p^{\cal K}(m_1, \cdots, m_N)$.
\end{observation}
This can be shown by expressing the correlation function as $C(t_{\mu_1}, \cdots, t_{\mu_j}, \cdots, t_{\mu_k}) = {\rm Tr}[ A(t_{\mu_{j+1}}) \cdots A(t_{\mu_k})\rho_{t_0} A(t_{\mu_1}) \cdots A(t_{\mu_j})]$, with two monotonically increasing sub-time sequences $t_{\mu_1} \leq t_{\mu_2} \leq \cdots \leq t_{\mu_j}$ and $t_{\mu_k} \leq t_{\mu_{k-1}} \leq \cdots \leq t_{\mu_{j+1}}$. The correlation function is then expressed in terms of $p^{\cal K}(m_1,\cdots,m_N)$ by noting that $\rho_{t_i} A$, $A\rho_{t_i}$, and $A \rho_{t_i} A$ can be simultaneously decomposed as a linear sum of ${\cal K}_{m_i} (\rho_{t_i})$ at each time $t_i$ (see Methods for detailed discussions).
We highlight that our approach allows a systematic protocol to obtain both time-ordered and out-of-time-ordered correlation functions from a single set of measurement data $p^{\cal K}(m_1, \cdots, m_N)$, without changing the setting for each correlation function.
For example, in the three-time case, one can additionally access the out-of-time-ordered correlation functions $C(t_3,t_2,t_1)$, $C(t_2,t_3,t_1)$, and $C(t_1,t_3,t_2)$.
The number of correlation functions that can be obtained from the $N$-time distribution $p^{\cal K}(m_1, \cdots, m_N)$ scales exponentially as $\approx 2^N$, since there are two choices for $A(t_i)$ to be placed either on the left or on the right sides of the quantum state $\rho_{t_i}$ at each time $t_i$. The out-of-time-ordered QPDs can also be obtained in the same way by taking $A(t_i) = \Pi_{x_i}(t_i)$.

As a special case, we can obtain the OTOC throughout quantum dynamics, which has been widely adopted as a quantifier of quantum information scrambling throughout complex quantum dynamics~\cite{Maldacena16, Landsman19} and has recently been studied in the context of QPDs~\cite{halpern2017jarzynski, halpern2018quasiprobability, Alonso2019OTOC}. The OTOC of a quantum system under unitary dynamics $U_\tau$ is defined as the absolute square of the commutator between two operators $V$ and $W$,
\begin{equation}
\label{eq:OTOC}
C_{\rm OTOC} \equiv \langle [W(\tau), V(0)]^\dagger [W(\tau), V(0)] \rangle,
\end{equation}
where $V(0) =V$ and $W(\tau) = U_\tau^\dagger W U_\tau$. We note that the OTOC is essentially a linear sum of four-point functions containing both time-ordered and out-of-time-ordered correlation functions. For example, if both $V$ and $W$ are Hermitian and unitary, $C_{\rm OTOC} = 2( 1- \langle W(\tau) V(0) W(\tau) V(0) \rangle)$.
Even though $C_{\rm OTOC}$ in Eq.~\eqref{eq:OTOC} contains terms with reversed time ordering, $p^{\cal K}(m_1, m_2, m_3)$ obtained from the three-time snapshotting method enables us to evaluate its value described as follows (see Methods):
\begin{observation}
\label{thm:otoc}
$C_{\rm OTOC}$ can be obtained from the sequential measurement outcomes $(m_1, m_2, m_3)$ at three-time points $(t_1, t_2, t_3)$ under the unitary dynamics $U_{t_1 \rightarrow t_2} = U_\tau$ and $U_{t_2 \rightarrow t_3} = U_\tau^\dagger$ as
\begin{equation}
C_{\rm OTOC} =\mathbb{E} \left[ \prod_{i=1}^3 \gamma_{m_i}^{\rm OTOC} \right],
\end{equation}
with the Kraus operator described in Theorem~\ref{thm:Kraus_decomp} and some complex coefficients $\gamma_{m_i}^{\rm OTOC}$.
\end{observation}
Compared to interference-based schemes for obtaining the OTOC~\cite{swingle2016measuring, zhu2016measurement, halpern2017jarzynski, yao2016interferometric, bohrdt2017scrambling}, our scheme offers the advantage that an ancilla state is required to remain coherent only for a short time during each ancilla-assisted measurement. We highlight that the time-reversal unitary is applied only once in our scheme. 
This can be contrasted with other ancilla-assisted measurement schemes~\cite{halpern2017jarzynski, halpern2018quasiprobability, dressel2018strengthening, mohseninia2019optimizing}, which possess the same advantage as our scheme in ancilla coherence time but require two time-reversals.
On the other hand, the stability of the protocol against imperfect implementations of the time-reversal unitary~\cite{Swingle2018resilience, Yoshida2019disentangling} remains open for quantitative comparison with an interference-based scheme without time reversals~\cite{yao2016interferometric, halpern2017jarzynski}. 
While a similar approach utilizing classical post-processing of generalized measurement outcomes was studied~\cite{dressel2018strengthening, mohseninia2019optimizing} for a qubit system under specific conditions $V^2 = \mathbb{1} = W^2$, our results hold for any diagonalizable operators $V$ and $W$ of a general $d$-dimensional system. Another important difference is that our scheme cancels the impact of the measurement at each time point via classical post-processing, whereas in Refs.~\cite{dressel2018strengthening, mohseninia2019optimizing}, the latter measurements undo the earlier measurements from the condition $V^2 = \mathbb{1} = W^2$.
We also note that both OTOC and QPD can be obtained from the same scheme in our approach without requiring subcircuits proposed in Ref.~\cite{mohseninia2019optimizing}.

\subsection{Experimental realization}
We experimentally demonstrate the proposed protocol with trapped ions. A crucial part of the protocol is the repeated measurement (ICD) and initialization (ICI) of the ancilla without influencing the system~\cite{kirchmair2009stateBlatt, schindler2011experimental,Home2018,ryan2021realization,decross2023qubit},
which are also core technologies for quantum error correction. In trapped-ion systems, the ICD and ICI can be achieved by adopting ion shuttling~\cite{kielpinski2002architecture, wan2019ion-shuntting, kaushal2020shuttling, pino2021demonstration, zhu2023interactive} or multiple types of qubits~\cite{home2013quantum, Tan&Wineland2015, Ballance2015, Inlek2017, Home2018, MIT2019, wang2021significant, allcock2021omg, yang2022realizing}. Here, we employ two different species trapped in a single trap, \Yb~and \Ba~ions~\cite{wang2017single, wang2021single, wang2021significant}, which are used for the system qubit and the ancilla qubit, respectively.
Both trapped ions are controlled by lasers with different wavelengths so that they can be controlled independently with minimal influence on each other~\cite{wang2021significant}.

In the experiment, the system qubit is encoded in the hyperfine levels of the $S_{1/2}$ manifold of the \Yb~ion, $\ket{F=0,m_F=0} = \ket{0}_\mathrm{Yb}$ and $\ket{F=1,m_F=0} = \ket{1}_\mathrm{Yb}$ with a splitting of 12.6428 GHz. The ancilla qubit is encoded in Zeeman levels of the $S_{1/2}$ manifold of the \Ba~ion, $\ket{s_j=1/2} = \ket{0}_\mathrm{Ba}$ and $\ket{s_j=-1/2} = \ket{1}_\mathrm{Ba}$ with an energy splitting of 16.2 MHz. Raman transitions are used to individually manipulate the \Yb~and \Ba~ion-qubits with 355 nm and 532 nm lasers, respectively. For the entangling operations for both qubits, we simultaneously apply the 355 nm and 532 nm laser beams with appropriately chosen frequencies (see Supplementary Fig.~1).

\begin{figure*}[htp!]
\centering
\includegraphics[width=180mm]{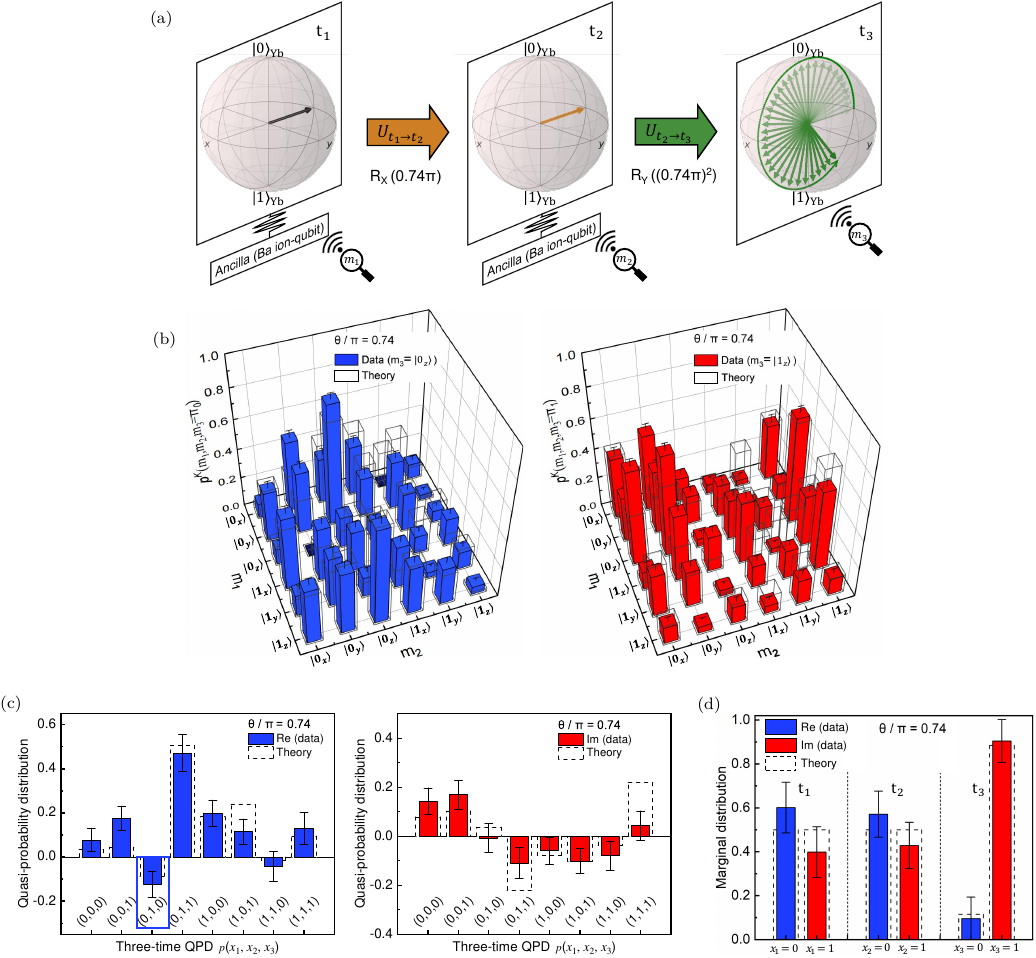}
\caption{Experimental reconstruction of the three-time QPD and marginal distribution by the ancilla-assisted measurement. 
(a) The system qubit's evolution is given by two successive rotations $U_{t_1 \rightarrow t_2} = R_X(\theta)$ and $U_{t_2 \rightarrow t_3} = R_Y(\theta^2)$ with $\theta = 0.74 \pi$.
Note that the measurements are performed at three time points $t_1$, $t_2$ and $t_3$.
(b) The normalized observed trajectories from the measurements at times $t_1$, $t_2$, and $t_3$ are represented by the 2D bar charts, where $m_1,m_2 \in \{ \ket{0_x}, \ket{0_y}, \ket{0}, \ket{1_x}, \ket{1_y}, \ket{1} \}$. The bar charts on the left and right represent the observed trajectories of $m_3 = \ket{0}$ and $m_3 = \ket{1}$, respectively, which are the measurement results of time $t_3$.
(c) The three-time QPD $p(x_1,x_2,x_3)$ reconstructed from the observed trajectories by classical processing. (i) The left blue bars indicate the real parts of the reconstructed three-time QPD, and the negativity of the real QPD is verified for $p(x_1 =0,x_2 =1,x_3 =0)$ (in the blue-line box), which is $-0.123(\pm 0.060)$. (ii) The right red bars indicate the imaginary parts of the reconstructed three-time QPD. 
(d) The marginal distributions for times $t_1$, $t_2$, and $t_3$ under the unitary dynamics $U_{t_1 \rightarrow t_2}$ and $U_{t_2 \rightarrow t_3}$ show the snapshotting of the state evolution. 
The distributions are marginalized over all the other time points of the QPDs.
For all figures, error bars indicate standard deviations (STDs).
}
\label{fig:wholescheme}
\end{figure*}

As a concrete example, we reconstruct the three-time QPDs following the quantum circuit of Fig.~\ref{fig:concept}(b). The initial state of the system qubit (\Yb~ion) is prepared to $\rho_{\rm Yb}=(\ket{1_x}\bra{1_x})_\mathrm{Yb}$, where $\ket{1_x}_\mathrm{Yb} = (\ket{0}_\mathrm{Yb} - \ket{1}_\mathrm{Yb}) /\sqrt{2}$ as described in Fig.~\ref{fig:wholescheme}(a).
At time $t_1$, the first ancilla-assisted measurement is performed. Then the system evolves from $t_1$ to $t_2$ under the unitary evolution $U_{t_1 \rightarrow t_2} = R_X(\theta) = e^{-i \frac{\theta}{2} X}$. At time $t_2$, the second ancilla-assisted measurement is conducted. 
From time $t_2$ to $t_3$, the system evolves under the unitary evolution $U_{t_2 \rightarrow t_3} = R_Y(\theta^2) = e^{-i \frac{\theta^2}{2} Y}$ and at time $t_3$, the system is directly measured (see Methods for further details).

The initial state and the time evolutions are chosen to illustrate how measurements can influence the subsequent dynamics of a quantum system and how the dynamics without the measurement effects can be reconstructed by our protocol discussed in the theory section. In particular, since the unitary evolution from time $t_1$ to $t_2$ is a rotation around the $x$-axis, the system state remains the same if no measurement is performed at time $t_1$. However, when a measurement is made on the $z$-basis, the state collapses to $\ket{0}$ or $\ket{1}$ and then rotates under the evolution $U_{t_1 \rightarrow t_2}$. 
From time $t_2$ to $t_3$, the unitary evolution rotates the system qubit around the $y$-axis with angle $\theta^2$, which leads to non-trivial behaviors of the QPDs and correlation functions beyond sinusoidal functions of $\theta$. 
The three-time QPD is then expressed as
\begin{equation}
\begin{aligned}
& p(x_1,x_2,x_3) \\
=& {\rm Tr}\left[ U_{t_2 \rightarrow t_3} U_{t_1 \rightarrow t_2} \rho_{t_1} \Pi_{x_1} U_{t_1 \rightarrow t_2}^\dagger \Pi_{x_2} U_{t_2 \rightarrow t_3}^\dagger \Pi_{x_3} \right]\\
=& \langle {1_x} \ket{x_1}\bra{x_1} e^{i (\theta/2) X} \ket{x_2}\bra{x_2} e^{i (\theta^2/2) Y} \ket{x_3}\bra{x_3}\\
& e^{-i (\theta^2/2) Y} e^{-i (\theta/2) X} \ket{1_x},
\end{aligned}
\end{equation}
where $\rho_{t_1} = \rho_{\rm Yb} = {(\ket{1_x}\bra{1_x}})_{\rm Yb}$ and $\Pi_{x} = \ket{x}\bra{x}$ with $x \in \{ 0,1 \}$ is an eigenprojector of $Z$.

\begin{figure*}[htp!]
\centering
\includegraphics[width=180mm]{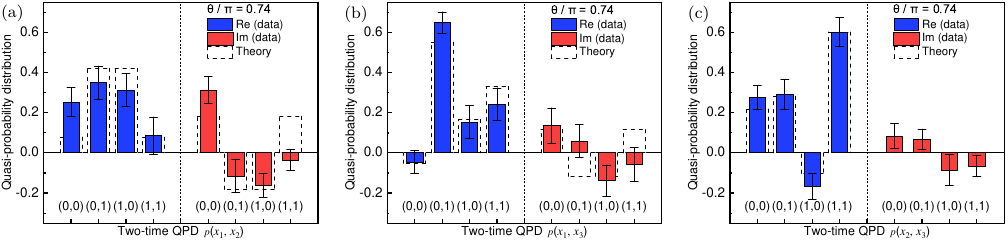}
\caption{The two-time QPDs obtained from the three-time QPDs. The blue and red bars indicate the real and imaginary parts of the QPDs. (a) The two-time QPD $p(x_1, x_2)$ from the unitary evolution $U_{t_1 \rightarrow t_2}$. (b) The two-time QPD $p(x_1, x_3)$ from $U_{t_1 \rightarrow t_3}$. (c) The two-time QPD $p(x_2, x_3)$ from $U_{t_2 \rightarrow t_3}$. For all figures, the error bars indicate the STD, and the theoretical expectations are shown as dashed bars (see Supplementary Note 4 for details). We note that all the results here are derived directly from the three-time QPD, rather than from a separate experiment.}
\label{fig:2pt-QPDs from 3pt}
\end{figure*}

Figure~\ref{fig:wholescheme} shows the experimental results for the above procedure.
The distribution of observed trajectories from three-time measurements, $p^{\cal K}(m_1,m_2,m_3)$, for the case of $\theta=0.74 \pi$ is shown in Fig.~\ref{fig:wholescheme}(b). 
At times $t_1$ and $t_2$, we have the measurement results in the $x$-, $y$- and $z$-basis and only the $z$-basis measurement results for time $t_3$.
As shown in the circuit of Fig.~\ref{fig:concept}(b), 
$x$-, $y$- and $z$-basis measurements of the ancilla yield six measurement outcomes, $m_i \in \{ \ket{0_x}, \ket{1_x}, \ket{0_y}, \ket{1_y}, \ket{0}, \ket{1} \}$.
We post-select the data with only dark state outcomes to avoid heating of the vibrational modes (see Supplementary Fig.~6 for further details). We repeat each measurement configuration 100 times, for a total of 3600 measurements.

From the distribution of observed trajectories in Fig.~\ref{fig:wholescheme}(b), the three-time QPDs $p(x_1,x_2,x_3)$ are reconstructed as shown in Fig.~\ref{fig:wholescheme}(c). The quasi-probability of $p(x_1=0, x_2=0, x_3=0)$, as an example, is obtained directly from the relation of Eq.~\eqref{eq:QPDs_from_traj}, $ \sum_{m_1, m_2, m_3} \gamma_{0 m_1} \gamma_{0 m_2} \gamma_{0 m_3} p^{\cal K}(m_1, m_2, m_3)$, where $\gamma_{0m}$ can be calculated from Eq.~\eqref{eq:qubit_decomp}. In our actual reconstruction, we perform an optimization procedure to obtain a proper $\gamma_{x_i m_i}$ for all experimental data (see Methods). 
Some data points in Fig.~\ref{fig:wholescheme}(c) deviate from the theoretical expectations by more than one standard deviation. This is because several observed trajectories shown in Fig.~\ref{fig:wholescheme}(b) deviate from the ideal values. However, these deviations are mainly due to technical imperfections rather than fundamental problems. We discuss experimental limitations related to fluctuations of experimental control parameters in the last section before the discussion section of the paper (see Supplementary Note 3 for further details).  
Despite these deviations, our experimental results reveal the essential features of the QPDs, which are different from classical probability distributions. Classically, the joint probabilities at multiple time points can only have positive values.
However, as shown in Fig.~\ref{fig:wholescheme}(c), the negative value for $p(x_1=0, x_2=1, x_3=0)$ and the imaginary values are observed for most cases.

The three-time QPDs enable us to evaluate the correct probability distribution of the system state during its unitary evolution.
By taking the marginals of the QPDs, we recover the probability distribution at each point in time, which is not influenced by the previous measurements. As shown in Fig.~\ref{fig:wholescheme}(d), the measurement results at times $t_1$, $t_2$, and $t_3$ are consistent with those distributions where no measurements were performed before. The clear difference with and without previous measurements can be seen in the probability distribution at time $t_3$. If projective measurements were performed at time $t_1$ or $t_2$, the distribution of the measurement results at time $t_3$ should be 0.5 for each basis, which is not the case as shown in Fig.~\ref{fig:wholescheme}(d). 

\begin{figure*}[htp!]
\centering
\includegraphics[width=180mm]{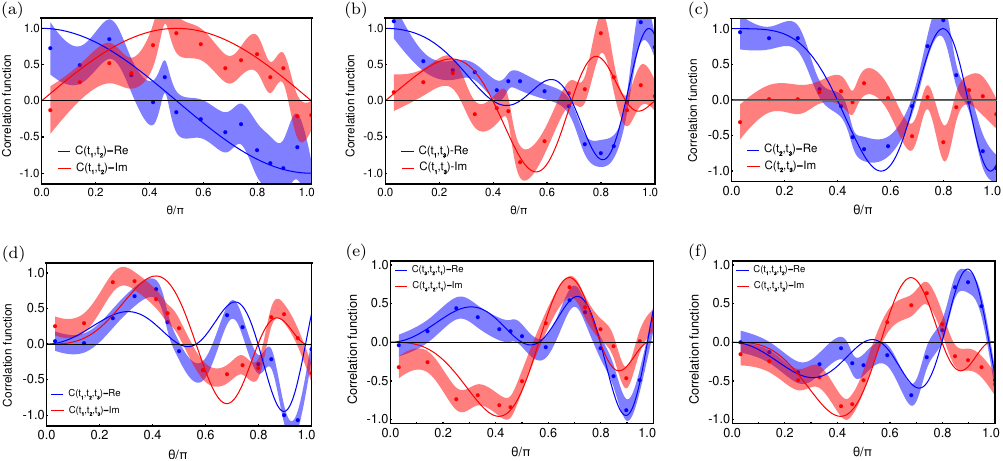}
\caption{The results of the two- and three-time correlation functions. Here, blue and red colors indicate the real and the imaginary parts of the correlation functions, respectively. Experimental data and theoretical expectations are shown as dots and solid lines, respectively. The bands indicate the STD of the experimental data.
(a-c) The two-time correlation functions reconstructed from the three-time QPDs. (d) The time-ordered three-time correlation function reconstructed from the three-time QPDs. (e-f) $C(t_3, t_2, t_1)$ with the reversed time-ordering and $C(t_1, t_3, t_2)$ with an increasing-decreasing time-ordering
respectively.}
\label{fig:correlation functions-2 and 3-pt and OTOC}
\end{figure*}

We can obtain any combination of two-time QPDs from the three-time QPDs and observe nonclassical features as shown in Fig.~\ref{fig:2pt-QPDs from 3pt}. The two-time QPDs are straightforwardly obtained from the three-time QPD by taking its marginals, $p(x_1, x_2) = \sum_{x_3} p(x_1, x_2, x_3)$, $p(x_1, x_3) = \sum_{x_2} p(x_1, x_2, x_3)$, and $p(x_2, x_3) = \sum_{x_1} p(x_1, x_2, x_3)$. We note that it is not always possible to do the reverse, that is, the reconstruction of the three-time QPDs from two-time QPDs, even with all possible combinations.
Some data points in Fig.~\ref{fig:2pt-QPDs from 3pt} deviate from the ideal values by more than their standard deviations.
Despite these deviations, we observe imaginary and negative values in two-time QPDs as shown in Fig.~\ref{fig:2pt-QPDs from 3pt}, which indicate the coherence of the state with respect to the measurement basis~\cite{Arvidsson-Shukur_2021, de2021complete, hance2023contextuality, Wagner2023simple}.

Imaginary and negative values in two-time QPDs are shown in Figs.~\ref{fig:2pt-QPDs from 3pt}(a-b) and Figs.~\ref{fig:2pt-QPDs from 3pt}(b-c), respectively. For the unitary evolution of the single-qubit state, the occurrence of imaginary and negative values in two-time QPDs can be understood from the relationship between the representation of the quantum state and the rotation axis in the Bloch sphere when the state contains coherence. For $U_{t_1 \rightarrow t_2}$, the initial quantum state is represented on the $x$-axis, and the rotation axis for the unitary evolution is also aligned along the $x$-axis in the Bloch sphere. In this case, the QPDs reveal imaginary values, as shown in Fig.~\ref{fig:2pt-QPDs from 3pt}(a). For $U_{t_2 \rightarrow t_3}$, the quantum state is on the $x$-axis of the Bloch sphere, but the axis of rotation is along the $y$-axis, perpendicular to the state. In this case, the QPDs reveal negative values, as shown in Fig.~\ref{fig:2pt-QPDs from 3pt}(c). Fig.~\ref{fig:2pt-QPDs from 3pt}(b) describing the process from time $t_1$ to $t_3$, which includes both parallel (time $t_1$ to $t_2$) and perpendicular (time $t_2$ to $t_3$) relationships between the initial state and the rotation axis.

We also obtain quantum correlation functions from the observed trajectories.
For different values of $\theta$, time-ordered three-time correlation functions (Fig.~\ref{fig:correlation functions-2 and 3-pt and OTOC}(d)) and all combinations of two-time correlation functions (Fig.~\ref{fig:correlation functions-2 and 3-pt and OTOC}(c)) are deduced from the three-time QPDs. We also reconstruct out-of-time-ordered correlation functions based on Observation~\ref{thm:corr_ft_out}.
We present $C(t_3, t_2, t_1)$ with the reversed time-ordering and $C(t_1, t_3, t_2)$ with an increasing-decreasing time-ordering in Figs.~\ref{fig:correlation functions-2 and 3-pt and OTOC}(e) and (f), respectively. The solid lines are from theoretical calculations, where their explicit forms can be found in Supplementary Note 4. 
Although certain data points exhibit deviations exceeding the error bars, the overall trends observed in the data are generally consistent with the theoretical predictions.

The majority of the experimental deviations stem from technical imperfections in controlling experimental parameters rather than fundamental issues in the underlying theoretical framework.
Fluctuations in the parameters of the M{\o}lmer-S{\o}rensen (M-S) gates are primarily responsible for the observed experimental deviations.
In particular, more than 90$\%$ of the data points that deviate from theoretical predictions in Fig.~\ref{fig:wholescheme}(b) can be explained by fluctuations in the rotation angle of the M-S gates and the relative phase between two successive gates (see Supplementary Note 3). In the analysis, the amounts of the fluctuations in rotation angle and relative phase are required to be 0.03$\pi$ (corresponding to 11.8$\%$ fluctuations) and 0.04$\pi$ (approximately 4.4$\%$ fluctuations), respectively.
We investigate the performance of the M-S gate using quantum process tomography, but we highlight that this data is not used for obtaining the joint distribution.
The related details and other experimental imperfections are discussed in Supplementary Note 3.

Fig.~\ref{fig_Marginal distribuction fidelity} indicates the fidelity $F(p,p_{\rm exp.}) = \sum_{x_3} \sqrt{p(x_3) p_{\rm exp.}(x_3)}$ between the theoretical distribution at time $t_3$ without intermediate measurements, $p(x_3) = {\rm Tr}[\rho_{t_3} \Pi_{x_3}]$, and the marginal distribution of the QPD, $p_{\rm exp.}(x_3) =\sum_{x_1,x_2}p_{\rm exp.}(x_1,x_2,x_3)$, obtained from the experimental data. The marginal distribution obtained experimentally at time $t_3$ is closer to the ideal distribution compared to the case when intermediate projective measurements are performed at times $t_1$ and $t_2$.

\begin{figure}[htbp]
\centering
\includegraphics[width=88mm]{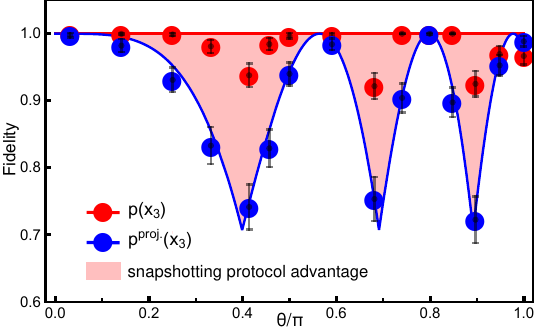}
\caption{
Fidelity between the distributions for the $z$-basis measurement at time $t_3$. The error bars represent the standard error of the mean. Red dots refer to the fidelity between the marginal distribution from theory ($p(x_3)$) and the experimentally obtained three-time QPD ($p_{\rm exp.}(x_3) = \sum_{x_1,x_2} p_{\rm exp.}(x_1,x_2,x_3)$). 
A fidelity of 1 refers to two distributions are equal. The blue line refers to the theoretical fidelity between $p(x_3)$ and $p^{\rm proj.}(x_3)$ when projective measurements are performed at times $t_1$ and $t_2$.
The blue dots refer to the corresponding experimental result. The pink shaded area indicates the advantage of the snapshotting protocol compared to projective measurements.}
\label{fig_Marginal distribuction fidelity}
\end{figure}

\section{Discussion}
We have introduced a protocol named snapshotting to extract quantum statistics at multiple times from ancilla-assisted measurements and demonstrated it experimentally using the \Yb-\Ba~trapped-ion system.
The key features of our approach are that the measurement effect can be entirely canceled out through classical post-processing of the ancilla measurement outcomes and that the measurement requires only a short-time system-ancilla interaction at each immediate time point. By snapshotting quantum dynamics, the QPD at multiple time points and various types of quantum correlation functions can be simultaneously obtained from a single distribution of observed trajectories.
We highlight that this method is applicable to any quantum system and dynamics, serving as a valuable experimental tool for exploring the quantum statistics of both open and closed quantum systems.

The potential applications of the proposed protocol include exploring quantum dynamics in many-body physics. In principle, when considering local observables, the number of samples required to reconstruct QPDs and correlation functions does not scale with the size of the system or the number of qubits. This property is promising for obtaining various critical quantities based on correlation functions, such as OTOC, in quantum many-body systems~\cite{Maldacena16, Landsman19}.
As the KD distribution itself has recently been recognized as an essential tool for investigating information scrambling and quantum thermodynamics~\cite{halpern2017jarzynski,halpern2018quasiprobability, Alonso2019OTOC, lostaglio2023kirkwood}, its direct reconstruction from experimental data will shed light on experimentally testing of the nonclassical phenomena arising from quantum dynamics.

\section{Methods}
\subsection{Proof of Theorem~\ref{thm:Kraus_decomp}}
Let us consider a slightly more general scenario such that the operators $A=\sum_{x=0}^{d-1} a_x \ket{x}\bra{x}$ and $B = \sum_{x=0}^{d-1}  b_x \ket{x}\bra{x}$ diagonal in the same basis $\{ \ket{x} \}_{x=0}^{d-1}$
are acting on the left and right sides of a $d$-dimensional quantum state, respectively, given as
\begin{equation}
{\cal E}_{B,A}(\rho) = B \rho A = \sum_m \gamma_{m}(B,A) {\cal K}_m(\rho),
\label{eq:Supp_BA}
\end{equation}
where ${\cal K}_m(\rho) = K_m \rho K_m^\dagger$. We note that $A$ and $B$ is not required to be Hermitian.
Theorem~\ref{thm:Kraus_decomp} in the main text to obtain the QPD $p(x_1, x_2, \cdots, x_N)$ is a special case with $A = \Pi_x = \ket{x}\bra{x}$, $B = \mathbb{1}$, and $\gamma_{xm} = \gamma_m(\mathbb{1}, \Pi_x)$.

We then construct a set of Kraus operators $\{ K_m \}$ to satisfy the condition in Eq.~\eqref{eq:Supp_BA}:
\begin{proposition}
\label{prop:meas}
For the operators $A = \sum_x a_x \ket{x}\bra{x}$ and $B = \sum_x b_x \ket{x}\bra{x}$ diagonal in the same basis $\{ \ket{x} \}$, there always exist $\gamma_m(B,A)$ satisfying Eq.~\eqref{eq:Supp_BA} for a set of Kraus operators $\{K_m\}$ with
\begin{equation}
K_m = \sum_{x=0}^{d-1} \left( \frac{\langle \phi_m | x \rangle}{\sqrt{\alpha}} \right)  \ket{x}\bra{x},
\end{equation}
where $\left\{ \ket{\phi_m}\bra{\phi_m} \right\}$ is a set of informationally complete projectors and satisfies $\sum_m \ket{\phi_m}\bra{\phi_m} = \alpha \mathbb{1}$.
\end{proposition}
\begin{proof}
\renewcommand{\qedsymbol}{} 
For the diagonal operators $A = \sum_x a_x \Pi_x$ and $B = \sum_x b_x \Pi_x$, let us rewrite Eq.~\eqref{eq:Supp_BA} as
\begin{equation}
\label{eq:suppl_proof1}
\begin{aligned}
    {\cal E}_{B,A}(\rho) &= B \rho A \\
&=  \left(\sum_{y=0}^{d-1} b_y \Pi_y \right) \rho \left( \sum_{x=0}^{d-1}  a_x \Pi_x \right)\\
    &= \sum_{x,y=0}^{d-1}  a_x b_y \Pi_y \rho \Pi_x\\
    &= \sum_{x,y=0}^{d-1} \langle x | \left( \sum_{x',y'=0}^{d-1} a_{x'}b_{y'} \ket{x'}\bra{y'} \right) | y\rangle \Pi_y \rho \Pi_x \\
    &= \sum_{x,y=0}^{d-1} \langle x | O(B,A) | y\rangle \Pi_y \rho \Pi_x,
\end{aligned}
\end{equation}
where we define $O(B,A) = \sum_{x,y =0}^{d-1} a_x b_y \ket{x}\bra{y}$. We note that
any operator $O$ can be expressed in terms of the informationally complete projectors $\{ \ket{\phi_m}\bra{\phi_m} \}$ as $O = \sum_m c_m^O \ket{\phi_m} \bra{\phi_m}$ with some complex coefficients $c_m^O$.  
This leads to an alternative expression
\begin{equation}
O(B,A) = \sum_m c_m^{O(B,A)} \ket{\phi_m}\bra{\phi_m}.
\end{equation}
By substituting this form into Eq.~\eqref{eq:suppl_proof1}, we obtain
\begin{equation}
    \begin{aligned}
        &{\cal E}_{B,A}(\rho)\\
        &= \sum_m \sum_{x,y=0}^{d-1} c_m^{O(B,A)} \langle x | \phi_m \rangle \langle  \phi_m | y\rangle \Pi_y \rho \Pi_x \\
        &= \sum_m \alpha c_m^{O(B,A)} \left( \sum_{y=0}^{d-1} \frac{\langle  \phi_m | y\rangle}{\sqrt\alpha} \Pi_y \right) \rho \left( \sum_{x=0}^{d-1} \Pi_x \frac{\langle x | \phi_m \rangle}{\sqrt\alpha}  \right)\\
        &= \sum_m \gamma_m(B,A) K_m \rho K_m^\dagger \\
        &= \sum_m \gamma_m(B,A) {\cal K}_m(\rho) ,
    \end{aligned}
\end{equation}
where we take $K_m = \sum_{y=0}^{d-1} \frac{\langle \phi_m  | y \rangle}{\sqrt\alpha} \Pi_y $ to satisfy the normalization condition $\sum_m K_m^\dagger K_m = \mathbb{1}$ and define $\gamma_m(B,A) = \alpha c_m^{O(B,A)}$.
\end{proof}

To complete the proof of Theorem~\ref{thm:Kraus_decomp}, we show that these Kraus operators can be realized by the ancilla-assisted measurements. 
To this end, we introduce a $d$-dimensional ancilla state, initially prepared in $\ket{0}_R$. After applying the CSUM gate, a generalized CNOT gate, 
$U_{\rm CSUM} = \sum_{x,y=0}^{d-1} \ket{x, x \oplus y} \bra{x, y}$
followed by the ancilla measurement 
with respect to the set of informationally complete projectors $\{ \ket{\phi_m} \bra{\phi_m}\}$, the Kraus operators for each measurement outcome become
\begin{equation}
K_m =  \sum_{x=0}^{d-1} \left( \frac{\langle \phi_m | x \rangle}{\sqrt{\alpha}} \right)  \Pi_x = \frac{\bra{\phi_m} U_{\rm CSUM} \ket{0}_R}{\sqrt{\alpha}}.
\label{eq:Supp_Mm}
\end{equation}
For a $d$-dimensional system, a set of informationally complete projectors $\{\ket{\phi_m}\bra{\phi_m}\}$ has at least $d^2$ elements.
For example, the measurement set discussed in the main text, 
$\{ \ket{\phi_m} \} = \{ \ket{0}, \ket{1}, \ket{0_y}, \ket{1_y}, \ket{0_x}, \ket{1_x} \}$ has $6$ elements, which is more than $d^2 = 4$, thus being overcomplete. 
However, this measurement set is easier to realize in experiments since all the projectors are the eigenvalues of the Pauli matrices.

Theorem~\ref{thm:Kraus_decomp} can be extended to local operators $A$ and $B$ of a multi-qudit system. This can be achieved by replacing $\Pi_x$ with $\mathbb{1} \otimes \cdots \otimes \mathbb{1} \otimes \Pi_x \otimes \mathbb{1} \otimes \cdots \otimes \mathbb{1}$, where the projection is only applied to the target qudit. Consequently, ${ K_m }$ only acts on the target qubit while maintaining the same form as in Eq.~\eqref{eq:Supp_Mm}, ensuring that the corresponding ancilla-assisted measurement requires only the interaction between the ancilla state and the target qudit. Since the coefficient $\gamma_m (B, A)$ for the local operators $A$ and $B$ remains the same as in the single-qudit case, the protocol does not scale with the size of the system as long as $A$ and $B$ act on a single-qudit.

We also note that there can be various choices of weight vectors $\gamma_m(B, A)$ that satisfy Eq.~\eqref{eq:Supp_BA} for a given set of Kraus operators $\{ K_m \}$.
In this case, the optimal choice would be to minimize $|\gamma(B,A)|_{\rm max} := \max_m \{ |\gamma_m(B, A)| \}$ 
as the number of samples to collect for a fixed precision scales with $|\gamma(B,A)|_{\rm max}^2$ from Hoeffding's inequality~\cite{Hoeffding63}.
More precisely, the optimization problem can be formalized as follows:
\begin{align}
    {\rm for~given:}&\qquad A,~B,~\{K_m\} \\
    {\rm minimize:}&\qquad |\gamma|_{\rm max} = \max_m \{ |\gamma_m| \}\\
    {\rm subject~to:}&\qquad B \otimes A^T = \sum_m \gamma_m K_m \otimes K_m^*.
    \label{eq:Suppl_subject_to}
\end{align}
By vectorizing the density matrix $\rho$ in Eq.~\eqref{eq:Supp_BA}, we note that Eq.~\eqref{eq:Suppl_subject_to} is equivalent to the condition that Eq.~\eqref{eq:Supp_BA} holds for any $\rho$.

For $A = \sum_{x=0}^{d-1} a_x \ket{x}\bra{x}$ and $B = \sum_{x=0}^{d-1} b_x \ket{x}\bra{x}$ and the measurement operators described in Eq.~\eqref{eq:Supp_Mm}, the condition in Eq.~\eqref{eq:Suppl_subject_to} is reduced to
\begin{equation}
    \boldsymbol{T} \boldsymbol{\gamma} = \alpha \boldsymbol{\xi},
\end{equation}
where $[\boldsymbol{T}]_{x+y d, m} = \langle \phi_m | y \rangle \langle x | \phi_m \rangle$, $[\boldsymbol{\gamma}]_m = \gamma_m$, and $[\boldsymbol{\xi}]_{x + yd} = a_x b_y$. 
From numerical optimization for the measurement set 
$ \{ \ket{\phi_m} \} = \{ \ket{0}, \ket{1}, \ket{0_y}, \ket{1_y}, \ket{0_x}, \ket{1_x} \}$ with $\alpha=3$, which leads to $\{ K_m\} = \left\{ \frac{\Pi_0}{\sqrt{3}}, \frac{\Pi_1}{\sqrt{3}}, \frac{S^\dagger}{\sqrt{6}}, \frac{S}{\sqrt{6}}, \frac{\mathbb{1}}{\sqrt{6}}, \frac{Z}{\sqrt{6}}  \right\}$, we obtain
$\gamma_{\rm max} = \max_{x,m} \{ | \gamma_m(\mathbb{1}, \Pi_x)| \}  | \approx 1.775$. 
This is a more efficient decomposition than that in Eq.~\eqref{eq:qubit_decomp} which yields $\gamma_{\rm max} = 3$.

\subsection{Derivation of Observation~\ref{thm:corr_ft_out}}
From the cyclic property of the trace, the correlation function can be rewritten as 
\begin{equation}
\begin{aligned}
&C(t_{\mu_1}, \cdots, t_{\mu_j}, \cdots, t_{\mu_k}) \\ 
&\quad=\Tr[ A(t_{\mu_{j+1}})\cdots A(t_{\mu_{k}}) \rho_{t_0} A(t_{\mu_1})\cdots A(t_{\mu_{j}})].
\end{aligned}
\end{equation}
Then we note that both $t_{\mu_1} \leq t_{\mu_2} \leq \cdots \leq t_{\mu_j}$ and $t_{\mu_k} \leq t_{\mu_{k-1}} \leq \cdots \leq t_{\mu_{j+1}}$ are monotonically increasing time sequences, which leads to the following expression:
\begin{equation}
\begin{aligned}
&C(t_{\mu_1}, \cdots, t_{\mu_j}, \cdots, t_{\mu_k}) \\ 
&\quad={\rm Tr}[({\cal E}_{B_N, A_N} \circ {\cal U}_{t_{N-1} \rightarrow t_N}  \circ  \cdots \circ {\cal U}_{t_1 \rightarrow t_2} \circ {\cal E}_{B_1, A_1})(\rho_{t_1})],
\end{aligned}
\end{equation}
where ${\cal U}_{t_i \rightarrow t_j}(\rho) = U_{t_i \rightarrow t_j} \rho U_{t_i \rightarrow t_j}^\dagger$ and we take $(B_i, A_i) = (\mathbb{1}, A)$ when $A(t_{i})$ is applied to the right side, $(B_i, A_i) = (A, \mathbb{1})$ when $A(t_{i})$ is applied to the left side, and $(B_i, A_i) = (A, A)$ when $A(t_{i})$ is applied to the both sides.

Since each ${\cal E}_{B_i, A_i}$ can be expressed as a linear combination of the actions of the Kraus operators $\{ K_m \}$ from Proposition~\ref{prop:meas}, we obtain the following form,
\begin{equation}
C(t_{\mu_1}, \cdots, t_{\mu_j}, \cdots, t_{\mu_k}) 
 = \mathbb{E} \left[ \prod_{i=1}^N \gamma_{m_i}(B_i,A_i) \right],
\end{equation}
by averaging over all possible observed trajectories following the distribution $p^{\cal K}(m_1, m_2, \cdots, m_N) = \Tr[ ({\cal K}_{m_N} \circ {\cal U}_{t_{N-1} \rightarrow t_N}\circ \cdots \circ {\cal U}_{t_1 \rightarrow t_2} \circ {\cal K}_{m_1}) (\rho_{t_1}) ]$.

We highlight that the correlation functions with different time sequences $(t_{\mu_1}, \cdots, t_{\mu_k})$ are obtained by only replacing the coefficients $\gamma_{m_i}(B_i, A_i)$, which can be easily done in classical post-processing using the same data used to obtain $p^{\cal K}$.

\subsection{Obtaining $C_{\rm OTOC}$ from intermediate measurements}
Let us express the OTOC for the two operators $W(\tau) = U_\tau^\dagger W U_\tau$ and $V(0) = V$ as
\begin{equation}
\begin{aligned}
& C_{\rm OTOC} = \langle [W(\tau), V(0)]^\dagger [W(\tau), V(0)] \rangle \\
&= \langle W^\dagger(\tau) V^\dagger(0) V(0) W(\tau) \rangle - \langle V^\dagger(0) W^\dagger(\tau) V(0) W(\tau)\rangle \\
&~ - \langle W^\dagger(\tau) V^\dagger(0) W(\tau) V(0) \rangle + \langle V^\dagger(0) W^\dagger(\tau) W(\tau) V(0)\rangle \\
&= {\rm Tr}[V U_\tau^\dagger W U_\tau \rho U_\tau^\dagger W^\dagger U_\tau V^\dagger ] \\
&~- {\rm Tr}[ V U_\tau^\dagger W U_\tau \rho V^\dagger U_\tau^\dagger W^\dagger U_\tau ]\\
&~- {\rm Tr}[ U_\tau^\dagger W U_\tau V \rho U_\tau^\dagger W^\dagger U_\tau V^\dagger ] + {\rm Tr}[ W U_\tau  V \rho V^\dagger U^\dagger_\tau W^\dagger] \\
&= 
\Tr[({\cal E}_{V,V^\dagger}  \circ {\cal U}^{-1} \circ {\cal E}_{W,W^\dagger} \circ {\cal U} \circ {\cal E}_{\mathbb{1}, \mathbb{1}})(\rho)]\\
&~ -\Tr[({\cal E}_{V, \mathbb{1}}  \circ {\cal U}^{-1} \circ {\cal E}_{W,W^\dagger} \circ {\cal U} \circ {\cal E}_{\mathbb{1},V^\dagger})(\rho)] \\
&~ - \Tr[({\cal E}_{\mathbb{1},V^\dagger}  \circ {\cal U}^{-1} \circ {\cal E}_{W,W^\dagger} \circ {\cal U} \circ {\cal E}_{V,\mathbb{1}})(\rho)]\\
&~ + \Tr[({\cal E}_{W,W} \circ {\cal U} \circ {\cal E}_{V,V^\dagger})(\rho)],
\end{aligned}
\end{equation}
where we denote ${\cal U}(\rho) = U_\tau \rho U_\tau^\dagger$ and ${\cal U}^{-1}(\rho) = U_\tau^\dagger \rho U_\tau$, respectively.

One can then construct the intermediate measurements $\{ K^V_{m_1} \} = \{ K^V_{m_3} \}$ and $\{ {\cal K}^W_{m_2} \}$ from Proposition~\ref{thm:Kraus_decomp} to express ${\cal E}_{\mathbb{1},\mathbb{1}}$, ${\cal E}_{\mathbb{1},V^\dagger}$, ${\cal E}_{V,\mathbb{1}}$, and ${\cal E}_{V,V^\dagger}$ as a weighted sum of ${\cal K}^V_{m_1}$ or ${\cal K}^V_{m_3}$, and ${\cal E}_{W,W^\dagger}$ as a weighted sum of ${\cal K}^W_{m_2}$. From this, all the four terms in $C_{\rm OTOC}$ can be obtained simultaneously from $p^{\cal K}(m_1, m_2, m_3) = \Tr[({\cal K}^V_{m_3}  \circ {\cal U}^{-1} \circ {\cal K}^W_{m_2} \circ {\cal U} \circ {\cal K}^V_{m_1}(\rho)]$.

\subsection{Experimental setup}
\begin{figure}[b]
 \centering
 \includegraphics[width=88mm]{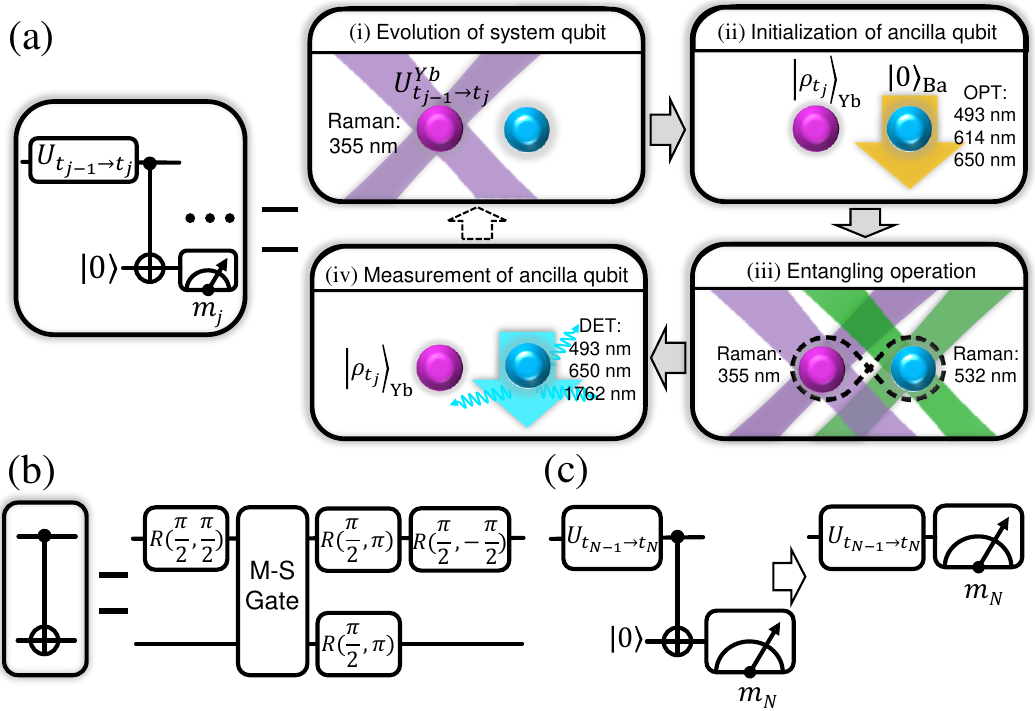}
 \caption{
 Experimental realization of unitary evolution and ancilla-assisted measurement with \Yb-\Ba~trapped-ion system. (a) For the step (i), the unitary operation $U_{t_{j-1} \rightarrow t_j}$ is performed by applying Raman laser beams to the system qubit represented by the pink ball. Later, the ancilla-assisted measurement is realized with the following three steps: (ii) initialization of the \Ba~qubit represented by the blue ball to $\ket{0}_\mathrm{Ba}$ by optical pumping (OPT), (iii) application of a CNOT gate between two qubits through an entangling operation, and (iv) measurement of the ancilla qubit with fluorescence detection (DET).
 (b) The CNOT gate consists of an M-S gate and four single-qubit rotations. The M-S gate can be described as $\exp(-i\frac{\pi}{4} X\otimes X)$, where $X$ is the Pauli operator. The single-qubit rotation is defined as $R(\theta,\phi)= \left(
  \begin{array}{cc}
    \cos(\frac{\theta}{2}) & -ie^{-i\phi}\sin(\frac{\theta}{2}) \\
    -ie^{i\phi}\sin(\frac{\theta}{2}) & \cos(\frac{\theta}{2})
  \end{array} \right)$.
  (c) The final time measurement can be performed by direct measurement in basis $m_{_N} \in \{\ket{0},\ket{1}\}$ on the system qubit.}
 \label{fig:Ancilla-assisted measurement}
 \end{figure}
We implement the quantum circuit in Fig.~\ref{fig:concept} to reconstruct the multi-time QPD for a qubit system. The experimental realization for the essential parts of the circuit is shown in Fig.~\ref{fig:Ancilla-assisted measurement}. At the beginning of the protocol, we optically pump the system qubit to $\ket{0}_\mathrm{Yb}$, then prepare the state of $\rho_{\rm Yb}$ by using a single-qubit rotation performed by applying 355 nm Raman laser beams. 
As depicted in Fig.~\ref{fig:Ancilla-assisted measurement}(a), the Raman lasers have a frequency difference that matches the transition frequency of the \Yb~ion-qubit. This frequency matching allows the Raman lasers to drive unitary evolutions, specifically single-qubit rotations, on the \Yb~ion-qubit.
At each time $t_1, \cdots, t_{N-1}$, we perform the ancilla-assisted measurement as illustrated in Fig.~\ref{fig:concept}(b) and Fig.~\ref{fig:Ancilla-assisted measurement}(a). The measurement procedure consists of initializing the ancilla qubit, applying a CNOT gate, and detecting the ancilla qubit state, where the first and third steps are regarded as the ICI and ICD. We perform the CNOT gate by using the M-S gate \cite{sorensen1999quantum} and single-qubit operations shown in Fig.~\ref{fig:Ancilla-assisted measurement}(b). The $z$-basis measurement of the \Ba~ion is realized by fluorescence detection after shelving $\ket{0}_\mathrm{Ba}$ to $D_{5/2}$ manifold. The $x$- and $y$-basis measurements are realized by rotating the axis of the \Ba~ion state before the $z$-basis measurement. 
To repeat the protocol, we take the following steps at each cycle: 
i) apply the unitary evolution to the system qubit,
ii) reset the ancilla qubit to realize the ICI,
iii) perform the ancilla-assisted measurement.
We simplify the final measurement by using a projection measurement on the system qubit of \Yb~ion in the basis $m_{_N} \in \{\ket{0}, \ket{1}\}$, since no further measurements are performed after that.

\subsection{Data availability}
The data that support the findings of this study are available from \href{https://doi.org/10.5281/zenodo.13736373}{Zenodo}.

\subsection{Code availability}
Code used in data analysis is available from \href{https://doi.org/10.5281/zenodo.13736373}{Zenodo}.

%

\subsection{Acknowledgments}
We thank the Innovation Program for Quantum Science and Technology under Grant (No.~2021ZD0301602 to K.K.). The National Natural Science Foundation of China Grants (No.~92065205 and No.~11974200 to K.K., No.~12304551 to P.W.). H.J.K. is supported by the KIAS Individual Grant No.CG085301 at the Korea Institute for Advanced Study. M.S.K. acknowledges the KIST Open Research Program.

\subsection*{Author contributions}
H.J.K. and M.S.K. proposed the protocol. P.W. and C.-Y.L developed the experimental system with the assistance of W.C, M.Q., Z.Z, K.W. P.W., and C.-Y.L implemented the protocol and led the data-taking. K.K. supervised the experiment. P.W., H.J.K., C.-Y.L, M.S.K., and K.K. wrote the manuscript.

\subsection*{Competing interests}
The authors declare that they have no competing interests.

\end{document}